\documentstyle[11pt,IAUS215,twoside,epsf]{article}

\def\edcomment#1{\iffalse\marginpar{\raggedright\sl#1\/}\else\relax\fi}
\reversemarginpar

\begin{document}
\vspace*{1cm}
\title{Rotating models for evolved low-mass stars}
 \author{Corinne Charbonnel}
\affil{LA-OMP (CNRS UMR 5572) 14, av.Belin, F-31400 Toulouse and Geneva Observatory, CH-1290 Sauverny}
\author{Ana Palacios}
\affil{LA-OMP (CNRS UMR 5572) 14, av.Belin, F-31400 Toulouse and \\
IAA Universit\'e Libre de Bruxelles, Campus de la Plaine, Bd du Triomphe, B-1050
Bruxelles}

\begin{abstract}
Low mass stars ($<$ 2-2.5M$_{\odot}$) exhibit, at all the stages of their 
evolution, signatures of processes that require challenging modeling 
beyond the standard stellar theory. 
In this paper we focus on their peculiarities while they climb the 
red giant branch (RGB). 
We first compare the classical predictions for abundance variations 
due to the first dredge-up with observational data in various environments.
We show how clear spectroscopic diagnostics probe the nucleosynthesis 
and the internal mixing mechanisms that drive RGB stars. 
Coherent data reveal in particular the existence of a non-standard mixing process 
that changes their surface abundances at the so-called RGB bump.
By reviewing the models presented so far to explain the various abundance 
anomalies, we show that the occurrence of this extra-mixing process is 
certainly related to rotation. 
Finally we discuss the so-called Li-flash which is expected to occur at the 
very beginning of the extra-mixing episode.
\end{abstract}

\section{Abundance anomalies in RGB stars due to in situ processes}

\subsection{Observational data}
According to the classical stellar evolution 
theory\footnote{By this we refer to the modeling of non-rotating, 
non-magnetic stars, in which convection is the only mixing process considered.}, 
the only opportunity for low mass stars (LMS) to modify their surface abundances  
happens on their way to the RGB when they do undergo the so-called first 
dredge-up (1DUP; Iben 1965). 
During this event their expanding convective envelope deepens in mass, 
leading to the dilution of the surface pristine material within regions 
that have undergone partial nuclear processing on the earlier main sequence. 
Qualitatively, this leads to the decrease of the surface abundances of the
fragile LiBeB elements and of $^{12}$C, while those of $^3$He, $^{13}$C and
$^{14}$N increase. 
Due to too low temperatures inside main sequence LMS, the 
abundances of O and heavier elements remain unchanged subsequent to the 1DUP.
Quantitatively, these abundance variations depend on the stellar mass 
and metallicity 
(e.g., Sweigart, Greggio \& Renzini, 1989; Charbonnel 1994; 
Boothroyd \& Sackmann 1999). 
After the 1DUP, the convective envelope withdraws while the hydrogen burning 
shell (HBS) moves outwards in mass. No more variations of the surface abundances 
are then expected until the star reaches the asymptotic giant branch.

The 1DUP predictions agree with the observations on the 
lower part of the RGB\footnote{One has of 
course to take into account possible variations of the surface Li 
abundance occurring already on the main sequence. 
This discussion is however out of the scope of the present paper 
(see Charbonnel, Deliyannis \& Pinsonneault 2000).}.
However observational evidences have accumulated, that we list below,  
of a second and distinct mixing episode that occurs in low mass
stars after the end of the 1DUP, and more precisely at the RGB bump. 

The determination of the carbon isotopic ratio ($^{12}$C/$^{13}$C, hereafter 
{\sl cir}) for RGB stars in open clusters with various turnoff masses 
(Gilroy 1989) provided the first pertinent clue on this process. 
It was indeed shown that bright RGB stars with initial masses lower 
than $\sim$ 2-2.5M$_{\odot}$ exhibit {\sl cir} considerably lower than 
predicted by the 1DUP.
Thanks to data in stars sampling the RGB of M67 (Gilroy \& Brown 1991), 
it clearly appeared that observations deviated from classical predictions just at 
the so-called RGB bump (Charbonnel 1994).
Field and globular cluster (GC) stars
behave similarly\footnote{The literature on the various abundance anomalies within GCs 
is so large that we chose to quote here only the most recent papers to illustrate 
our discussion. See Palacios (2002) for complete references.}. 
In the former case this could be established thanks to the determination of 
the {\sl cir} in large samples of stars for which Hipparcos parallaxes 
allowed the precise determination of their evolutionary status 
(Charbonnel, Brown \& Wallerstein 1998; Gratton et al. 2000). 
We knew for a long time that the brightest RGB stars in GCs 
presented {\sl cir} close to the equilibrium value. 
But it is only this year that the region around the bump 
could be probed for two GCs : In NGC 6528 and M4, the {\sl cir} drops 
below the 1DUP predictions just at the RGB bump (Shetrone 2002). 
The disagreement between observations and classical predictions is stronger
for lower stellar mass and metallicity. 

The extra-mixing process affects the surface abundances of other chemical elements : 
Li also decreases at the RGB bump, both in the field and in GCs 
(Pilachowski, Sneden \& Booth 1993; Grundahl et al. 2002). 
In field stars, C decreases while N increases for RGB stars brighter than the bump 
(Gratton et al. 2000) confirming the CN processing of the stellar envelope. 
The lowering of the C abundance along the RGB is also seen in GCs 
(e.g., Bellman et al. 2001 and references therein), though in this case 
the picture is more confused because of a probable non negligible 
dispersion of the initial [C/Fe]. 

An in situ mechanism has also frequently been invoked to explain the 
abundance anomalies of heavier elements in GC stars, and in particular
the O-Na anticorrelation.  
This pattern has been observed in the brightest GC RGB stars for a long time 
(see references in Ivans et al. 1999 and Ramirez \& Cohen 2002). 
However it is only thanks to 8-10m-class telescopes that O and Na abundances 
could be determined for less evolved stars, and in particular for turnoff stars 
in a couple of GCs (Gratton et al. 2001; Th\'evenin et al. 2001; Ramirez \& Cohen 2002). 
There, the O-Na anticorrelation extends to the main sequence. 
This result is crucial. Indeed, the ON- and NeNa-cycles do not occur in 
main sequence LMS, as the involved reactions require too high temperatures 
which are only reached when LMS are on the RGB. 
Thus the fact that the O-Na anticorrelation already exists on the main sequence 
clearly proves that it is not produced by in situ processes, but 
by external causes, the discussion of which is out of the scope of this paper 
(see Charbonnel 2002b).

Recently we have searched for an eventual evolution of the O and Na abundances with luminosity 
along the RGB for field stars. 
We gathered from the literature a large sample of field (625) stars with Hipparcos 
parallaxes over a large range of metallicity ([Fe/H] between -2.5 and 0) 
for which we homogeneously redetermined the abundances of a large number of elements. 
No evolutionary effect could be found neither for O nor Na 
(Palacios et al. 2002a; Charbonnel et al. 2003), in agreement with the
most recent GC data mentioned above.

Last but not least, the recent determination of the oxygen isotopic ratios
in a couple of RGB stars with low {\sl cir} (Balachandran \& Carr 2002) 
confirmed this result.  
In these objects indeed, both the $^{16}$O/$^{17}$O and $^{16}$O/$^{18}$O ratios 
appear to be high, in agreement with extensive CN-processing but no dredge-up of 
ON-cycle material in RGB stars (see the arrow in Fig.1).

\subsection{Origin of the extra-mixing and its inhibition by $\mu$-gradients}
The observations summarized previously provide definitive clues on a second and 
distinct mixing episode that occurs in LMS when they reach the RGB bump.
This process appears to be universal : 
it affects more than 95$\%$ of the LMS (Charbonnel \& do Nascimento 1998), 
whether they belong to the field, to open or globular clusters.
Its signatures in terms of abundance anomalies are clear: 
the Li and the $^{12}$C abundances as well as the $^{12}$C/$^{13}$C ratio drop, 
while the $^{14}$N abundance and the $^{16}$O/$^{18}$O ratio increase. 
On the other hand $^{16}$O, $^{17}$O, Na and heavier elements remain unaffected.
As we shall see in \S 3, this sequence may start 
with a very brief episode of Li enrichment. 

These coherent data clearly indicate that during the extra-mixing episode at the RGB bump,
the stellar convective envelope is connected with the radiative regions 
where CN-burning occurs, but not with the ON- and NeNa-processed layers (see Fig.1).
As we shall discuss later on, the important increase of the molecular weight 
(or $\mu$) gradients when one approaches the HBS certainly shields 
these deeper regions from the extra-mixing. 

\begin{figure}
\plotfiddle{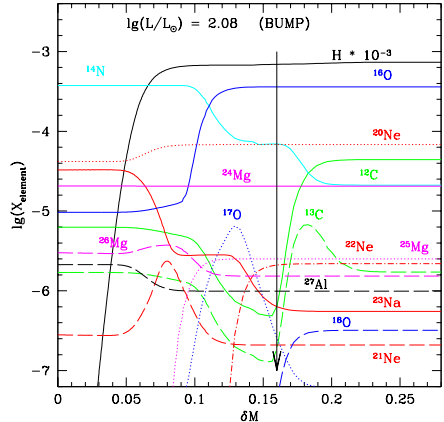}{5cm}{0}{30}{30}{-100}{-50}
\caption{Abundance profiles of the isotopes participating to the CNO, NeNa and MgAl
cycles inside a 0.83M$_{\odot}$, [Fe/H]= -1.6 (typical of the GC M13) standard model 
at the luminosity of the RGB bump. 
The absissa is the relative mass coordinate 
$\delta M =$(m$_r$-m$_{BHBS}$)/(m$_{BEC}$-m$_{BHBS}$) 
($\delta M = 0$ at the bottom of the HBS, $\delta M = 1$ at the base of the convective 
envelope). 
The vertical arrow indicates the maximum depth down to which the extra-mixing process 
is efficient according to the observations}
\end{figure}

$\mu$-gradients are also clearly responsible for the fact that the 
extra-mixing starts to be efficient only around the bump. 
Indeed during the 1DUP a $\mu$-discontinuity is built at the region of the 
deepest penetration of the convective envelope. This $\mu$-barrier then 
inhibits any mixing between the convective envelope and the HBS. 
At the RGB bump, the external regions of the HBS pass through the 
$\mu$-discontinuity. Due to the composition changes in the burning region,
the stellar structure has to readjust, and the stellar luminosity then 
slightly decreases before increasing again. 
LMS stars spend a non negligible part of their RGB lifetime in this region; 
this explains why, for GCs for example, the bump clearly appears as a peak 
in the differential luminosity function or as a change in the slope of the
cumulative luminosity function (e.g., Zoccali et al. 1999). 
After this evolutionary point,
the $\mu$-gradients between the base of the convective envelope and the HBS 
are much smoother, allowing the occurrence of some extra-mixing in this radiative region 
(Sweigart \& Mengel 1979; Charbonnel 1994, 1995; Charbonnel et al. 1998).

All these observational facts represent strong clues on the origin of the 
extra-mixing process which is certainly rotationally-induced.

\section{Rotation-induced mixing and abundance anomalies in RGB 
stars\footnote{We do not discuss here the numerous parametric models that 
have been constructed to reproduce the abundance anomalies in RGB stars. 
We focus only on the physical models where the extra-mixing is assumed 
to be induced by rotation.}}
\subsection{From meridional circulation ...}
Sweigart \& Mengel (1979, hereafter SM79) investigated the possibility that 
meridional circulation might lead to the mixing 
of CNO-processed material in RGB stars.
Though the physics of the rotation-induced mixing invoked was incomplete, 
this pioneering work has magnificently settled the basis of a complex problem 
which is not definitively solved yet. 

SM79 discussed in great details the problem of $\mu$-gradients which 
were known to tend to inhibit meridional circulation (Mestel 1953, 1957). 
For the mixing to be efficient, they underlined the necessity for the radiative 
zone separating the convective envelope from the HBS not to have a significant 
gradient in $\mu$. 
As we said in \S 1.2, this condition is fulfilled only after the RGB bump, 
and, above this evolutionary point, only in the most external part of the HBS. 
This last point can be easily understood by looking at the abundance profiles
inside a RGB star at the bump (Fig.1). When one moves inward the star from the
base of the convective envelope ($\delta$M=1) toward the HBS (X$_{\rm H}$=0 
at $\delta$M=0), one encounters
first a shell in which C is burned into N, then a C-depleted region and still 
deeper a shell in which O burns into N. As underlined by SM79, the H abundance,
and thus $\mu$, remain essentially constant across the C shell :
the negligible $\mu$-gradients can not chocke the 
circulation currents in this region. However, the $\mu$-gradients become appreciable when one 
approaches the O shell due to the greater amount of H burning close to ON 
equilibrium. There, it is no longer clear whether the circulation currents
can overcome the $\mu$-gradients, especially for the more metal-rich stars 
in which the C and O shells are shifted inward toward the HBS.
Theoretical estimates of the $\mu$-gradients which are required to stop 
the meridional circulation were too uncertain for definitive conclusions to be put forward, 
though the metallicity dependance was soundly addressed.

The other crucial point made by SM79 concerned the importance of the 
angular momentum distribution within the deep convective envelope of the RGB
star on the resulting CNO processing.
Indeed, the angular velocity 
of a radiative layer near the HBS depends on how much angular momentum is 
left behind as the convective envelope retreats outward. 
SM79 investigated the two extreme cases, namely a constant specific angular momentum 
within the convective envelope, and a convective envelope rotating as a solid body.
Substantial CNO processing of the envelope could be obtained with plausible 
main sequence angular velocity only if the inner part of the convective envelope
was allowed to depart from solid body rotation. 
As we shall see in \S 2.3, our lack of knowledge of the distribution of angular momentum 
within the RGB convective envelope still remains nowadays one of the weakest points 
of the global picture. 

\subsection{... over a more complete description of rotation-induced mixing}
Rotational transport processes can not be simply reduced to meridional circulation.
Once established indeed, this large scale circulation generates in turn 
advection of momentum and thus favours the development of various hydrodynamical 
instabilities\footnote{We refer to the review by Suzanne Talon in these proceedings 
for a complete description of the transport processes inside rotating stars}. 
Zahn (1992) proposed a description of the interaction between meridional circulation and 
shear turbulence, 
pushing forward the idea of chocking the meridional circulation 
by $\mu$-gradients. 
Following these developments, Charbonnel (1995, hereafter C95) re-investigated 
the influence of such a process in RGB stars. 
This framework is appealing because it takes advantage of some particularities
of the non-homologous RGB evolution.
In particular, some mixing is expected to take place wherever the rotation
profile presents steep vertical gradients, as well as near nuclear burning shells.
Moreover due to the stabilizing effect of the composition gradients,
the mixing is expected to change the surface abundances on the RGB only after the bump.

Using a simplified version of Zahn's description, C95 showed that
the rotation-induced mixing can indeed account for the observed behavior
of carbon isotopic ratios and for the Li abundances in Population II 
low mass giants.  
Simultaneously, when this extra-mixing begins to act, $^3$He is
rapidly transported down to the regions where it burns by the
$^3$He($\alpha, \gamma)^7$Be reaction. This leads to a decrease
of the surface value of $^3$He/H and of the final $^3$He yields from LMS, 
solving the long-standing problem of the galactic evolution of $^3$He 
(e.g. Charbonnel 2002a).
In these exploratory computations however, the transport of angular momentum 
by the hydrodynamical processes was not treated self-consistently, although 
it is of utmost importance in the understanding of the rotation-induced mixing.

\subsection{ ... to a self-consistent treatment of the transport of angular 
momentum and of the chemicals}
Maeder \& Meynet (2000) nicely review the recent developments on 
the rotation-induced mixing in stars. They clearly state the fact that 
``the study of rotational mixing requires both an understanding of 
angular momentum evolution of stars and an understanding of the degree 
of mixing for a given angular momentum distribution. 
Angular momentum transport is the crucial ingredient for determining 
the extent of rotationally induced mixing". 
Let us now discuss the results obtained for RGB stars 
when the transport of angular momentum and of the chemicals as 
described by Zahn (1992), and latter on by Talon et al. (1997), 
Maeder \& Zahn (1998), and Palacios et al. (2002b) is taken into account.

Two independant studies have been carried out using basically the 
same physical description of rotational mixing : 
Denissenkov \& Tout (2000, hereafter DT00) and Palacios et al. (2003, hereafter PCTF03).
They differ on some details of the physical assumptions like the choice of the 
shear instability criteria and of the expression of 
the diffusion coefficient associated with vertical turbulence.  
These differences actually appear to be minor. 
Both studies obtain indeed very similar total diffusion coefficients 
for Pop II RGB stars at the bump, the vertical diffusion coefficient (turbulence) 
dominating over the effective diffusion coefficient (meridional circulation).

More importantly, however, a drastic assumption differentiates the work by 
DT00 from that of PCTF03 : DT00 indeed compute the transport of 
angular momentum and of the chemicals in a post-processing way; namely, 
they use the internal structure of a few {\sl standard} RGB models to 
estimate the transport coefficients and the subsequent expected changes 
in the surface abundances. 
In this approach, there is no feedback of the mixing on the stellar 
structure and evolution which are computed independently in the standard way.

On the other hand, PCTF03 do treat the transport of both angular momentum 
and of the chemicals inside their stellar evolution code, thereby following 
the ``non-standard" evolution of the star from the zero age 
main sequence on. 
As a result in this case, the mixing has a feedback on the star since its early 
evolution. On the RGB in particular, the mixing distorts the stationary 
profiles of the chemicals deep inside the star even if the surface 
abundances are not yet affected; it alters for example the positions 
of the C and O depleted regions.  
In turn, the mixed star is allowed to react to the mixing that it 
undergoes, thereby influencing the mixing itself.

This difference has a strong influence on the final predictions. 
In DT00 indeed, the transport is applied on standard profiles, and 
surface variations of O and Na are expected in RGB stars.
In PCTF03 however, the mixing is applied on self-consistently distorted
profiles; as a result, only variations of the C isotopes and of N
are predicted at the surface of the star.
These latter predictions are in better agreement with the observations 
described in \S 1.


\subsection{Clues from horizontal branch stars}
Let us note an important point. In all these studies (SM79, DT00, PCTF03), 
notable changes of the surface abundances could be reached  
only under the assumption of constant specific angular momentum within the
convective envelope (see \S 2.1). Indeed in the case of solid body rotation 
prohibitively high values of the main sequence surface rotation velocities
are required to produce enough mixing. 
The same result has been obtained by Chanam\'e, Pinsonneault \& Terndrup 
(these proceedings) in their simplified rotating Pop I RGB model.

Unfortunately, the distribution of angular momentum within the very deep 
convective envelope of an RGB star is far from being understood.
More than 20 years after SM79, one is still reduced to use the abundance anomalies 
on the RGB as evidences for some departure from solid-body rotation 
(see also Kumar, Narayan \& Loeb 1995).
This assumption is actually sustained by the study of horizontal branch stars 
which provides some intriguing clues about angular momentum evolution on the RGB.  

Peterson (1983) first discovered that some blue horizontal branch stars 
are rapid rotators. Pinsonneault et al. (1991) noted that
the combination of RGB mass loss, high horizontal branch rotation rate, and
low main sequence rotation required strong differential rotation with
depth in giants.  If the convection zone of RGB stars had solid body
rotation, differential rotation with depth in their MS precursors was
also required.
Behr et al. (2000) found a break in the rotational properties of horizontal branch 
stars, in the sense that very blue horizontal branch stars both exhibited the surface
signature of atomic diffusion and rotated more slowly than slightly
cooler stars.  Sills \& Pinsonneault (2000) interpreted this as
an evidence that mean molecular weight gradients caused by atomic
diffusion inhibit angular momentum transport in hot horizontal branch
stars. This is an independent test of the impact of composition
gradients in a different evolutionary phase 
(for discussions of the effects on the main sequence see for example Vauclair 1999; 
Th\'eado \& Vauclair 2001; Palacios et al. 2002b).

In addition, Sills \& Pinsonneault (2000) found that uniform rotation 
at the main sequence turnoff was only compatible with rapid 
horizontal branch rotation under the following conditions:
(1) turnoff rotation of order 4 km/s rather than the 1 km/s inferred
from an extrapolation of the Population I angular momentum loss law to
Population II stars;
(2) constant specific angular momentum in the convective envelopes of
giants;
(3) strong differential rotation with depth in the radiative cores of
giants.

All of these are different from the expectations from main
sequence angular momentum evolution models, and they are an indication
that further theoretical work is needed in physical models of RGB
rotational mixing.  It is encouraging, however, that all of the above
properties favor more vigorous rotational mixing on the RGB than would
be expected from the opposite conclusions.

\section{The Li flash}
A few ($\sim 1\%$) RGB stars present Li-overabundances 
(see also Drake et al. and Konstantinova-Antova in these proceedings). 
The fact that these so-called super Li-rich giants are all located at the RGB bump
lead Charbonnel \& Balachandran (2000, hereafter CB00) to conclude that the Li-rich phase was 
a precursor to the extra-mixing process described previously. 
This brought very tight constraints on the underlying physics.  

Denissenkov \& Weiss (2000, hereafter DW00) investigated the effect of rotation-induced 
mixing on the surface lithium abundance using the transport coefficients D$_{\rm R}$
derived by DT00 (see \S 2.3). 
They found that the corresponding values of D$_{\rm R}$ 
($10^8$ to 10$^{10}$cm$^2$sec$^{-1}$) were too low
to lead to lithium enrichment and invoked the engulfing of a planet in order to
trigger and dope the transport process. While this combined scenario is
rather attractive, it is harmed by the fact that it can happen at any time
on the RGB while the super-Li rich giants with solar metallicity discovered 
up to now do stand by the bump (CB00).
However what was not investigated by DT00 and DW00 was the structural
response of the star to the extra-mixing. Indeed, as already discussed in 
\S 2.3, their computations of the transport coefficients and of the 
abundance variations were done in a post-processing approach 
where standard stellar structures were used as background models.

Palacios, Charbonnel \& Forestini (2001) proposed that 
the structural response to the mixing could actually be 
the cause of the increase of D$_{\rm R}$ which is necessary to
produce super-Li rich giants.
The mixing sequence is the following.
At the RGB bump the external wing of the $^7$Be peak crosses the
molecular weight discontinuity before the $^{13}$C peak does, and 
is then connected with the convective envelope by the extra-mixing process. 
$^7$Be produced via $^3$He($\alpha,\gamma)^7$Be starts to diffuse outwards. 
However, due to the relatively low D$_{\rm R}$ given above, the transported 
$^7$Be decays in regions where $^7$Li is rapidly destroyed by proton capture.
A lithium burning shell appears.
$^7$Li(p,$\alpha)\alpha$ becomes the dominant reaction leading to an increase of the
local temperature and of both the local and total luminosities.
Palacios et al. suggested that the meridional circulation and the corresponding 
transport coefficient then increase due to the $\epsilon_{\rm nuc}(^7$Li+p) burst, 
allowing for lithium enrichment of the convective envelope.
A lithium flash terminates this phase\footnote{Our scenario is different from 
the previous works which explored the
possible occurrence of shell flashes on the RGB
and which considered only the reactions of the CNO cycle
(Bolton \& Eggleton 1973; Dearborn et al. 1975; Von Rudloff et al. 1988).
It also differs from the model developed by Fujimoto et al. (1999)
to explain Mg and Al anomalies in globular cluster red giants.
These authors indeed proposed that a flash is triggered off deeper in the HBS
due to the inward mixing of H down into the degenerate He core.}.
As the convective instability develops around the Li-burning shell, 
it erases the molecular weight gradient and the mixing is free 
to proceed deeper. Both the lithium abundance and the
carbon isotopic ratio then drop at the stellar surface. 
When the flash is quenched, the external convective envelope deepens again.
We suggest that this creates a new $\mu$-barrier, similar to that built up
during the 1DUP, which then inhibits again the extra-mixing during the 
subsequent evolution. 

During the whole sequence described above the temporary increase of the stellar luminosity
causes an enhanced mass loss rate which naturally accounts for the dust shell suggested
by the far-infrared color excesses measured for some lithium-rich objects from
IRAS fluxes (de la Reza et al. 1996, 1997).
Let us note that the contribution of these stars to the lithium enrichment of the Galaxy
should be very modest.

\section{Conclusions}
During the last three decades, an incredible amount of work has been devoted 
to the understanding of the chemical anomalies exhibited by evolved LMS. 
On the observational side, crucial breakthroughs were made possible recently 
thanks to the advent of 8-10m-class telescopes. 
The signatures of the extra-mixing process that occurs at the RGB bump are 
now very clear. 

On the other hand, recent theoretical models include a sophisticated description
of the mixing processes induced by rotation. Those which consistently couple the
transport of angular momentum and of the chemicals with the structural evolution of the 
star predict surface abundance variations on the RGB in agreement with the observational data.
However, a couple of assumptions remain to be tested. One of the main 
uncertainties concerns the distribution of angular momentum within the deep 
convective envelope. A final answer to this problem certainly necessitates 
3D hydrodynamical simulations. 
Another important point which remains to be investigated is the reaction of the 
meridional circulation and of the various instabilities to a major and local release 
of nuclear energy.  

Rotation-induced mixing processes modify all the stellar outputs, 
and in particular the final yields from LMS 
(i.e., $^3$He is not over-produced anymore by rotating LMS). 
Their impacts on the stellar lifetimes and on the further evolution (luminosity 
of the RGB tip, morphology of the horizontal branch, AGB phase, ...) have to be 
investigated now in a systematic way. 
This is of crucial importance if one wants to correctly understand 
the role of LMS in the evolution (chemical and spectro-photometric) of stellar 
clusters and of galaxies in general. 

\acknowledgments We thank the french Programme National de Physique Stellaire and 
Programme National Galaxies for their support on this work. AP aknowledges financial
support from the IAU.

\end{document}